\documentclass[epsf,graphics,a4,12pt]{article}
\usepackage{epsfig}
\begin{document}
\baselineskip=18pt
\newcommand{\be}{\begin{equation}}
\newcommand{\en}{\end{equation}}
\begin{titlepage}
\begin{center}
{\hbox to \hsize{\hfill .}}
\bigskip
\vspace{6\baselineskip}
{\Large \bf Holography and the Shannon's First Theorem.}
\end{center}
\vspace{1ex}
\centerline{\large
L.Alejandro Correa-Borbonet\footnote{correa@fma.if.usp.br}}
\begin{center}
{Instituto de F\'{\i}sica, Universidade de S\~{a}o Paulo,\\
C.P.66.318, CEP 05315-970, S\~{a}o Paulo, Brazil}
\vspace{4ex}
\large{

{\bf Abstract}\\
}
\end{center}
\noindent
A possible link between Holography and Information Theory is
presented. Using the relation between the Shannon and Boltzmann
formulas Holography can be seen as the best encoding scheme. 
\\
\\
\vspace{9ex}
\hspace*{0mm} PACS number(s): 04., 89.70.+c
\vfill
\end{titlepage}

\newpage
\section{Introduction.}
The phrase "Information is physical", enunciated by Rolf Landauer
\cite{landauer} is the core of the research field known as physics of
information. The author claimed that information is everywhere, and, whether it is in biological systems, in a computer, or in a black
hole, inevitably has a physical form.  Also information, of whatever
kind, is associated with matter, radiation or fields of different type. It
is impossible to get away from this sort of physical
embodiment. Therefore, manipulation of information is inevitably
subject to the laws of physics. 

On the other hand there is a particular physical conjecture known as the
Holographic Principle\cite{susskind}, which states that the number of the 
degrees of freedom describing the physics inside a volume (including
gravity) is bounded by the area of the boundary which encloses this volume.

As information is physical we can also try the other way around,
interpreting physical theories in terms of the language of the Information
Theory. More precisely, in this paper I will attempt to connect the
holography with one of the cornerstone of Information Theory, the
Shannon noiseless coding theorem\cite{shannon}.

\section{Information Theory .} \nonumber
The classical theory of information is due to a Bell Labs mathematician, 
Claude Shannon, who in two seminal works definitively laid down its
principles. The works of Shannon deal with two very important tasks of
information processing :(i)file compression 
and (ii)error correction. The Shannon noiseless coding theorem -also called source coding theorem- addresses the first issue: How much redundancy can be eliminated from a given source text without losing information? The Shannon noisy coding theorem addresses the second issue: How much redundancy must be added to a given source text in order to guarantee error free communication over a noisy channel.

In this paper we will concentrate just on the first theorem and its relation with holography.

\subsection{Definitions and example}
Information is discretized: it comes in irreducible packages. The
elementary unit of classical information is the bit(or cbit, for
classic bit), a classical system with only two states, $0$ and
$1$. Any text can be coded into a string of bits\cite{roman},
\cite{galindo}.

{\bf Definitions}
\begin{itemize}
\item Let $S :=\{s_1,s_2,...,s_n\}$ be a finite alphabet, endowed
with a probability distribution $P:s_{j}\rightarrow p_{s}(s_{j})$
\item Codeword - a string of bits which represents a symbol in the
alphabet.
\item Code, codebook - the set of all possible codewords, e.g., the
4-bit binary code has 16 code-words
\item Let $\Gamma=(S,P)$ be a source. An encoding scheme for $\Gamma$
is an ordered pair $(C,f)$, where $C$ is a code and $f:S\rightarrow C$
is an injective function, called encoding function.
\end{itemize}
For the purposes of noiseless encoding, the measure of efficiency of
an enco\-ding scheme is its {\it average codeword length}.

{\bf Definition:} The {\it average codeword length} of an encoding
scheme $(C,f)$ for a source $\Gamma=(S,P)$  is defined by
\be
AveLen(C,f)=\sum_{i=1}^{N} p_{i} \, len(f(s_{i}))
\en 
where $len(f(s_{i}))$ is the length of the string that represents
$s_{i}$.

The Noiseless Coding Theorem, first proved by Claude Shannon in 1948, asserts 
that the {\it average codeword length} of any instantaneous code is bounded from below by $H(C)$, i.e.$H(C)\leq AveLen(C)$, where $H$ is the so called Shannon's entropy
\be
H(C)=-\sum_{i=1}^{N} p_{i}  log_{2}  p_{i} \label{eq:eshannon}
\en
The randomness of an information source can be describe by its
entropy. The operational meaning of entropy is that it determines the
smallest number of bits per symbol that is required to represent the
total output. 

{\bf Example}

Consider an alphabet $\{a,b,c,d\}$ with probabilities
$\{ \frac{1}{2},\frac{1}{4},\frac{1}{8},\frac{1}{8}\}$. With four
alternatives, it seems natural to assign two-bit binary codewords to
each symbol, eg, $\{00,01,10,11\}$. If this coding scheme is used, the
average codeword length is
\be
AveLen(l)=\frac{1}{2} 2+\frac{1}{4} 2
+\frac{1}{8} 2+\frac{1}{8} 2 =2.
\en 
However,  we can do better than this. The entropy of
this alphabet is
\be
H=-\left(\frac{1}{2}log_{2}\frac{1}{2}+\frac{1}{4}log_{2}\frac{1}{4}+\frac{1}{8}log_{2}\frac{1}{8}+\frac{1}{8}log_{2}\frac{1}{8}\right)=\frac{7}{4}
\en  
and Shannon's theorem says that a code exists which achieves this
rate. In this case, it is easy. The code $\{0,10,110,111 \}$ assigns
shorter codewords to the more frequent symbols. The average codeword
length is
\be
AveLen(l)=\frac{1}{2} 1+\frac{1}{4} 2+\frac{1}{8} 3
+\frac{1}{8} 3 =\frac{7}{4} .
\en
In the simple example above, the Shannon entropy and the average code
length were both $7/4$ bits/symbol. It happened this way because the
pro\-babilities were powers of $1/2$. We are not so fortunate in reality
and there will always be a difference between the Shannon's entropy
and the achieved average code length. This is quantified by two
numbers, the ${\it efficiency}$ and the ${\it redundancy}$
\be
\mathrm{Efficiency} =\frac{H}{AveLen}\leq 1
\en
\be
\mathrm{Redundancy} =1-\mathrm{Efficiency}.
\en
\section{Holography}
Physicists have the standard vision that the degrees of freedom of
the world consist of fields filling space. Improving this scenario some
theorists proposed that a small distance cutoff will be required in
order to make sense of quantum gravity. Therefore according to that
hypothesis the world could be seen as three dimensional discrete lattice with
spacing order of the Plank length, $l_{p}$. But recently, 't Hooft and
Susskind\cite{susskind} suggested an idea even more audacious. According to them the
combination of Quantum Mechanics and Gravity requires the three
dimensional world to be an image of data that can be stored on a two
dimensional projection much like a holographic image. This description
only requires one discrete degree of freedom per Planck area and yet
it is rich enough to describe all three dimensional
phenomena. Therefore it is obvious that the assumption of these ideas
imply a radical decrease in the number of degrees of freedom for
describing the Universe.

The inspiration for the holographic principle comes from the black
hole physics where the entropy of a black hole is given by the
Bekenstein-Hawking formula\cite{beken},\cite{hawk}
\be
S=\frac{A}{4G}=\frac{A}{l^{2}_{p}}log 2,
\en
where A is the horizon area. This result was obtained when they
realized the striking resemblance between the laws of
black hole mechanics and the laws of thermodynamics. Later, a great
amount of work has been done in order to find a precise statistical
mechanical interpretation of black hole entropy. One could derive the
Bekenstein-Hawking formula by counting black hole microstates and
using the Boltzmann entropy formula
\be
S=ln\; W,
\en
where $W$ is the number of microstates of the system.  

 This counting is not a simple task because is quite difficult to
identified the black hole microstates. However great progress in this
direction has been reported\cite{vafa}.  

Therefore,  the basic idea of holography is that the entropy
scales like area, or in other words, the total number of states is of
order
\be
W_{A} \sim e^{A}.
\en

This is a very dramatic change of view because, from
the quantum field theory experience, generally one expects that if the
energy density is bounded then the maximum entropy is proportional to
the volume of space and this means
\be
W_{V} \sim e^{V}.
\en

Therefore, for our purposes, the most important conclusion that we can
extract from the holographic principle is that the radical reduction
of the degrees of freedom necessary for the description of the universe
can be expressed in the following simple way,
\be
W_{V} \rightarrow W_{A}.\label{eq:redu}
\en

\section{Holography as the best Code}
The tools and concepts developed in the above sections will help to
show the main idea of this work. Thus, the language of Information
Theory will be used to deal with Holography.

Firstly we should point out the relation between the Boltzmann and
Shannon formulas. Despite the fact that they had different origins
they are conceptually equivalent\cite{jaynes} and the  formal relation is very
simple,\footnote{The relation can be obtained when in
(\ref{eq:eshannon}) the outcomes are equally likely, that is, when
$p_{i}=\frac{1}{W}$.}
\be
S=\frac{1}{log_{2} e} \; H. \label{eq:boshann}
\en
Here we ignore the units of the thermodynamical entropy and we
consider just the bit units.
 
This relation allows to establish the proper bridge between
the two main topics of this work.

Now we arrive to our point. First we consider a gravitational system
with a  corresponding `physical alphabet` ${\bf F}$.  This `alphabet`
can be thought as the source of the dynamic of the system. The next
step is to consider the codes.

Let us suppose that  there exits an `holographic code' ${\bf A}$ for this gravitational
system and the corresponding average codeword length is
\be
AveLen({\bf A})=\frac{A}{4}log_{2} e.
\en
Similarly, if instead of that we take a `volumetric code'${\bf V}$, we assume that
\be
AveLen({\bf V})=V \, log_{2} e.
\en
On the other hand, if the Holographic Principle is satisfied, the
Boltzmann entropy is $S=A/4$. Therefore, using (\ref{eq:boshann}) we get
\be
H=\frac{A}{4}log_{2} e.
\en
Obtaining,
\be
H=AveLen({\bf A}).
\en
Consequently, with the support of  the Noiseless Coding Theorem we can
conclude that the 'holografic code' is the best encoding scheme that
can be constructed.

In this context a volumetric quantum field theory can be seen as a
highly inefficient encoding scheme where  
\be
\mathrm{Efficiency} =\frac{A}{V} \sim 1/R \ll 1.
\en

\section{Conclusions}

Fermat  pointed out that nature is economical, and that light will
therefore travel along the path that takes the least time.  This is
quite a good rule but there are some instances where it is erroneous 
and the modern statement is extended: the ray path along which
light travels is such that the time taken holds a stationary value (i.e. 
minimum, maximum or constant).  Extending this argument to physics
of information we could speculate that Holography is a manifestation of
the economical character of nature. The minimum amount of information
is used to describe the Universe.

However, that is not the end of the story. We can go on further in the
same direction studying also the possible connections of Holography
with the Shannon's Second Theorem. Also the arguments presented here
could be refined using the Quantum Source Coding Theorem, introduced
by Schumacher\cite{schum}

Taking into account that the main idea of this work is encoded on the
title I will finish with the hope that I introduced the less possible redundancy.

{\bf ACKNOWLEDGMENT:}
The author is grateful to Elcio Abdalla for reading the manuscript and
useful suggestions. This work was supported
by Funda\c{c}\~ao de Amparo \`a Pesquisa do Estado de
S\~ao Paulo (FAPESP).


\end{document}